\documentstyle[12pt]{article}

\begin{document}

\font\fortssbx=cmssbx10 scaled \magstep2
\hbox to \hsize{
\hskip.5in \raise.1in\hbox{\fortssbx University of Wisconsin - Madison}
\hfill$\vcenter{\hbox{\bf MADPH-96-924}
            \hbox{January 1996}}$ }
\vskip 2cm
\begin{center}
\Large
{\bf Semi-leptonic $B$ decays into higher charmed resonances} \\
\vskip 0.5cm
\large
  Sini\v{s}a Veseli  and M. G. Olsson  \\
\vskip 0.1cm
{\small \em Department of Physics, University of Wisconsin, Madison,
	\rm WI 53706}
\end{center}
\thispagestyle{empty}
\vskip 0.7cm

\begin{abstract}
We apply HQET to semi-leptonic $B$ meson decays into
a variety of excited charm states. Using three realistic meson models
with fermionic light degrees of freedom, we examine the extent
that the sum of exclusive single charmed states account for the
inclusive semi-leptonic $B$ decay rate. The consistency of
form factors with the Bjorken and Voloshin
sum rules is also investigated.
\end{abstract}

\newpage
\section{Introduction}

The role of semi-leptonic $B$-decay in determining the CKM matrix element
is well known. An important ingredient in this role is the understanding
of the hadronic form factor, or Isgur-Wise (IW) function \cite{isgur}. For
$B\rightarrow D^{**}e \bar{\nu}_{e}$ decays the heavy-light limit is an
excellent starting point.\footnote{We use the symbol $D^{**}$ to denote any
charmed meson.} Heavy quark effective theory (HQET) \cite{reviews}
is useful in organizing the descriptions of such processes. In this limit
the wave functions and energies of the light degrees of freedom (LDF)
are sufficient to define all IW functions \cite{modelling}.

As the number of measured $B$ decays increases, other questions related to
semi-leptonic decay can be addressed. Among these are the $B\rightarrow D^{*}$
spectrum shape, and the ratio of $D^{*}$ to $D$ rates. Of particular
interest here are the inelastic processes $B\rightarrow D^{**}Xe
\bar{\nu}_{e}$,
where $D^{**}$ could be $D_{1}$ or $D_{2}^{*}$ (or their
radial excitations), and $X$ are any non-charmed hadrons. Interest
in these inelastic decays stems partly from experimental data
\cite{cleo,aleph,opal} for exclusive inelastic decays, and partly
from the relation between the exclusive decay rates and measured and
theoretical results for the inclusive semi-leptonic rate.

An additional motivation for considering inelastic exclusive channels
concerns sum rules. The Bjorken   sum rule \cite{bjsum,bjsum2} relates
the slope of the elastic IW function (at the zero recoil point) to
inelastic IW form factors describing $S$ to $P$-wave
semi-leptonic $B$ decays.
A less known sum rule involving the $S$ to $P$-wave
form factors  has been derived
by Voloshin  \cite{voloshin}. It is the analog of the ``optical''
sum rule for the dipole scattering of light in atomic physics.
It is not immediately
obvious whether a particular model yields form factors
which are consistent with both of these two sum rules.

Since the explicit IW functions are defined in terms of a relativistic
LDF wave functions with a fermionic light quark, we have
considered three appropriate hadronic models, based on the Dirac or the
Salpeter equations. The models considered are
the Dirac equation with scalar confinement, the Salpeter
equation with time component vector confinement, and
a relativistic flux tube model. In each case the strong coupling
constant and the various quark masses are
chosen to best fit the heavy-light data. The resulting
form factor predictions are quite similar and the variety of
different models used provides an assessment of the confidence one has
in predictions of this sort.

In Section \ref{ff} we briefly review our previous exclusive form factor
results, and in Section \ref{hl} we outline the three Dirac-like
hadronic models that are used for our numerical results. In Section \ref{res}
predictions for elastic and inelastic
branching ratios into a single charmed hadron are compared with other
results that can be found in the literature.
A discussion of
 theoretical and experimental results of fractional semi-leptonic decay rates
 is found in Section \ref{res2}. By considering fractional
 inclusive rates it is plausible that much of the uncertainty associated
with the $V_{cb}$ value, quark masses, and QCD corrections cancels out.
The consistency of form factors with
the Bjorken and Voloshin sum rules is considered in Section \ref{bj}.
A summary of the results and our conclusions are given in Section \ref{con}.

\section{Isgur-Wise form factors and semi-leptonic $B$ decays in the
heavy-quark limit}
\label{ff}

In the  $m_{Q}\rightarrow \infty$ limit
the angular momentum  of the LDF  decouples from the spin of the heavy quark,
and both
are separately conserved by the strong interaction. Therefore, total
angular momentum $j$ of the LDF  is a good quantum number. For
each $j$ there are two
degenerate heavy meson states ($J=j\pm \frac{1}{2}$), and we
can label states as $J^{P}_{j}$.

In HQET
the covariant trace formalism \cite{georgi2,korner,falk}
 is the most convenient
for keeping track
of the relevant Clebsch-Gordan coefficients and
for counting  the number of independent form factors.
Using the notation of \cite{mannel},
the lowest lying mesonic states are labeled as follows:
$C$ and $C^{*}$ denote $0^{-}_{\frac{1}{2}}$ and
$1^{-}_{\frac{1}{2}}$ states ($L=0$ doublet),
($E$, $E^{*}$) and  ($F$, $F^{*}$) denote the two
$L=1$ doublets
 ($0^{+}_{\frac{1}{2}}$, $1^{+}_{\frac{1}{2}}$) and
($1^{+}_{\frac{3}{2}}$, $2^{+}_{\frac{3}{2}}$), respectively, and
$G$ and $G^{*}$ denote
$1^{-}_{\frac{3}{2}}$ and
$2^{-}_{\frac{3}{2}}$ states ($L=2$ doublet).
These states are described by $4\times 4$ matrices,
and matrix elements of the heavy quark currents are calculated by
taking the corresponding traces  \cite{falk}. In this way, one
can write the decay rate for $B\rightarrow D^{**}e \bar{\nu}_{e}$ in the form
\begin{equation}
\frac{d\Gamma^{**}}{d\omega} = \frac{G_{F}^{2}|V_{cb}|^{2}}{48\pi^{3}}
m_{B}^{2}m_{D^{**}}^{3}\sqrt{\omega^{2}-1}|\xi^{**}(\omega)|^{2}
f^{**}(\omega,r^{**})\ .
\label{width}
\end{equation}
Here $\omega=v\cdot v'$ denotes velocity transfer,
 $r^{**}=m_{D^{**}}/m_{B}$, and the function $f^{**}$
is given by
\begin{eqnarray}
f_{C}(\omega,r_{C}) &\hspace*{-2mm}=\hspace*{-2mm}&
(\omega^{2}-1)(1+r_{C})^{2}\ ,\\
f_{C^{*}}(\omega,r_{C^{*}}) &\hspace*{-2mm}=\hspace*{-2mm}&
(\omega+1)[(\omega+1)(1-r_{C^{*}})^{2} +
4\omega(1-2\omega r_{C^{*}} + r_{C^{*}}^{2})]\ ,\\
f_{ E}(\omega,r_{ E}) &\hspace*{-2mm}=\hspace*{-2mm}&
(\omega^{2}-1)(1-r_{ E})^{2}\ ,\\
f_{E^{*}}(\omega,r_{E^{*}}) &\hspace*{-2mm}=\hspace*{-2mm}&
(\omega-1)[(\omega-1)(1+r_{E^{*}})^{2} +
4\omega(1-2\omega r_{E^{*}} + r_{E^{*}}^{2})]\ ,\\
f_{F}(\omega,r_{F}) &\hspace*{-2mm}=\hspace*{-2mm}&
\frac{2}{3}(\omega-1)(\omega+1)^{2}
[(\omega-1)(1+r_{F})^{2} +
\omega(1-2\omega r_{F} + r_{F}^{2})]\ ,\\
f_{F^{*}}(\omega,r_{F^{*}}) &\hspace*{-2mm}=\hspace*{-2mm}&
\frac{2}{3}(\omega-1)(\omega+1)^{2}
[(\omega+1)(1-r_{F^{*}})^{2} +
3\omega(1-2\omega r_{F^{*}} + r_{F^{*}}^{2})]\ ,\\
f_{G}(\omega,r_{G}) &\hspace*{-2mm}=\hspace*{-2mm}&
\frac{2}{3}(\omega-1)^{2}(\omega+1)
[(\omega+1)(1-r_{G})^{2} +
\omega(1-2\omega r_{G} + r_{G}^{2})]\ ,\\
f_{G^{*}}(\omega,r_{G^{*}}) &\hspace*{-2mm}=\hspace*{-2mm}&
\frac{2}{3}(\omega-1)^{2}(\omega+1)
[(\omega-1)(1+r_{G^{*}})^{2} +
3\omega(1-2\omega r_{G^{*}} + r_{G^{*}}^{2})]\ .
\end{eqnarray}
The above expressions can be found in
\cite{neubert}-\cite{stop}.

The only unknown quantity in the expression (\ref{width}) is the
appropriate IW form factor $\xi^{**}(\omega)$. Since these form factors
 cannot be calculated
from  first principles, one has to rely on some model of
strong interactions. By comparing the wave function approach of
\cite{zalewski} with the covariant trace formalism of
\cite{georgi2,korner,falk}, and performing the necessary integrations
in the modified Breit frame (as suggested in \cite{sadzikowski}), one finds
the expressions for the unknown form factors in terms of the
LDF rest frame wave functions and energies \cite{modelling}.
These expressions  include transitions from the ground state
into radially
 excited states, and
are given as
\begin{eqnarray}
\xi_{C}(\omega)& =&
\frac{2}{\omega+1}
\langle j_{0}(ar)\rangle_{00}
\hspace*{+2.65cm} (C\rightarrow C,C^{*}\ {\rm transitions})
\ ,\label{xic}\\
\xi_{E}(\omega)& =&
\frac{2}{\sqrt{\omega^{2}-1}}
\langle j_{1}(ar)\rangle_{10}\label{xie}
\hspace*{+2.08cm} (C\rightarrow E,E^{*}\ {\rm transitions})\ ,\\
\xi_{F}(\omega)& =&
\sqrt{\frac{3}{\omega^{2}-1}}\frac{2}{\omega+1}
\langle j_{1}(ar)\rangle_{10} \label{xif}
\hspace*{+1.05cm}(C\rightarrow F,F^{*}\ {\rm transitions})\ ,\\
\xi_{G}(\omega)& =&
\frac{2\sqrt{3}}{\omega^{2}-1}
\langle j_{2}(ar)\rangle_{20}
\hspace*{+2.45cm} (C\rightarrow G,G^{*}\ {\rm transitions})\ .
\label{xig}
\end{eqnarray}
In the above formulae $a$ is defined in terms of the initial ($E_{q}$)
and final ($E'_{q}$) energies of the LDF as
\begin{equation}
a = (E_{q}+E'_{q})\sqrt{\frac{\omega-1}{\omega+1}}\ .
\end{equation}
The expectation values appearing in (\ref{xic})-(\ref{xig}) are defined
as
\begin{equation}
\langle F(r) \rangle_{j'j}^{\alpha'\alpha} = \int r^{2} dr
[f^{*k'}_{\alpha' j'}(r)f^{k}_{\alpha j}(r) +
g^{*k'}_{\alpha' j'}(r)g^{k}_{\alpha j}(r) ] F(r)\ .
\end{equation}
This follows from the form of the wave function
in the Dirac-like models with spherical symmetry,
\begin{equation}
\phi_{j\lambda_{j}}^{(\alpha k)}({\bf x}) =
\left( \begin{array}{c}
f_{\alpha j}^{ k}(r) {\cal Y}_{j\lambda_{j}}^{k}(\Omega)\\
i g_{\alpha j}^{ k}(r) {\cal Y}_{j\lambda_{j}}^{-k}(\Omega)\end{array}
\right)\ ,
\label{wfd}
\end{equation}
where ${\cal Y}_{j\lambda_{j}}^{k}$ are the usual
spherical spinors,
 $k=l$ ($l=j+\frac{1}{2}$) or $k=-l-1$ $(l=j-\frac{1}{2})$, and
$\alpha$  denotes all other quantum numbers.

\section{Modelling the light degrees of freedom}
\label{hl}

In order to obtain a reasonable estimate of the model
dependence of our results, we employ  three qualitatively different
realistic models to describe  heavy-light mesons: the Dirac equation
with scalar confinement (DESC), the Salpeter equation with vector
confinement (SEVC), and the relativistic flux tube confinement (RFTC).
All three models involve a short range Coulomb potential.
With wave function of the form (\ref{wfd}), it can be shown
\cite{observations,fermionic} that
all three models satisfy a radial equation of the form
\begin{equation}
(E_{q} 1\hspace{-1.65mm}1 - I\hspace{-1.65mm}H_{0} - I\hspace{-1.65mm}L
I\hspace{-1.65mm}I I\hspace{-1.65mm}L)
\left(\begin{array}{c}
f_{j}^{k}(r) \\
g_{j}^{-k}(r) \end{array}
\right) = 0\ ,
\end{equation}
where
\begin{equation}
I\hspace{-1.65mm}H_{0} = \left(
\begin{array}{cc}
m_{q} & -D_{-} \\
D_{+} & -m_{q}
\end{array}
\right) \ ,
\end{equation}
and
\begin{equation}
D_{\pm}
= \pm \frac{k}{r} + (\frac{\partial}{\partial r} + \frac{1}{r})\ .
\end{equation}
The $2\times 2$ matrices $I\hspace{-1.65mm}L$ and $I\hspace{-1.65mm}I$ will be
defined when the
specific models are discussed. The numerical methods used to deal
with these three models are described in \cite{observations,
fermionic}.

\subsection{Dirac equation with scalar confinement (DESC)}

Scalar confinement is the only type of confinement potential
that has correct sign of the spin-orbit coupling. In the Dirac equation
it also yields linear Regge trajectories.
This model also assumes a
time component vector short range Coulomb interaction. Specifically, we have
\begin{eqnarray}
I\hspace{-1.65mm}L &=& \left(
\begin{array}{cc}
1 & 0\\
0 & 1
\end{array}
\right) \ , \\
I\hspace{-1.65mm}I &=& \left(
\begin{array}{cc}
-\frac{4\alpha_{s}}{3r} + b r & 0\\
0 & -\frac{4\alpha_{s}}{3r} - b r
\end{array}
\right) \ .
\end{eqnarray}
The parameter values chosen to give an excellent fit (see Table \ref{tall})
to the heavy-light spin averaged data \cite{pdg} are
\begin{eqnarray}
m_{u,d}&=&0.300 GeV\ \ \ {\rm (fixed)}\ ,\nonumber \\
m_{s} &=& 0.465 GeV\ ,\nonumber \\
m_{c}&=&1.357 GeV\ ,\nonumber \\
m_{b} &=& 4.693 GeV\ ,
\label{fit_desc}\\
\alpha_{s} &=& 0.462\ ,\nonumber\\
b&=&0.284 GeV^{2}\ \ \ {\rm (fixed)}\ .\nonumber
\end{eqnarray}
The quality of fit is insensitive to the value of $m_{u,d}$ and the
confinement tension $b$ was chosen to yield the universal
Regge slope \cite{stop}.\footnote{The  slope of the Regge trajectories
 in the
heavy-light case is expected to
be  exactly twice the slope  in the light-light case
\cite{observations,goebel},
i.e. $\alpha'_{HL} = 2\alpha'_{LL}$.  The observed
Regge slope for the light-light states is $\alpha'_{LL} = 0.88\ GeV^{-2}$
\cite{barger}.}

\subsection{Salpeter equation with vector confinement (SEVC)}

The instantaneous version of the Bethe-Salpeter
equation \cite{bethe,gellmann}
(usually referred to as the Salpeter equation \cite{salpeter})
is widely used for the discussion of bound state problems.
It is also equivalent \cite{long}
to the so called ``no-pair'' equation \cite{sucher}, which
was
introduced  in order to avoid the problem of mixing of
positive and negative energy states that occurred in the
Dirac equation for the helium atom. A similar
problem also occurs for a single fermionic particle moving in the
confining Lorentz vector potential. For a very long time \cite{plesset}
it has been known that there are no normalizable solutions
to the Dirac equation in this case.

It has been shown
 analytically  for the heavy-light case \cite{observations},
and numerically for the case of fermion and antifermion with
arbitrary mass \cite{lagae,instantaneous}, that in this type
of model linear scalar confinement does not yield
linear Regge trajectories.
We have therefore used time component
vector confinement with short range Coulomb interaction, even though
it is well known that this model
 gives the wrong sign of the spin-orbit coupling.

In this model matrices $I\hspace{-1.65mm}L$ and $I\hspace{-1.65mm}I$ are given
by
 \begin{eqnarray}
I\hspace{-1.65mm}L &=& \left(
\begin{array}{cc}
\lambda_{+} & -\frac{1}{2E_{0}^{+}}D_{-}\\
D_{+}\frac{1}{2E_{0}^{+}} & \lambda_{-}
\end{array}
\right) \ , \label{llnpvc}\\
I\hspace{-1.65mm}I &=& \left(
\begin{array}{cc}
-\frac{4\alpha_{s}}{3r} + b r & 0\\
0 & -\frac{4\alpha_{s}}{3r} + b r
\end{array}
\right) \ ,
\end{eqnarray}
where
\begin{equation}
\lambda_{\pm} = \frac{E_{0}^{\pm}\pm m_{q}}{2 E_{0}^{\pm }}\ ,
\end{equation}
and
\begin{equation}
E_{0}^{\pm} = \sqrt{m^{2}_{q}-D_{\mp}D_{\pm}}\ .
\end{equation}
Again we fix the confinement tension $b$ to yield the
universal Regge slope, and choose our parameters as
\begin{eqnarray}
m_{u,d}&=&0.300 GeV\ \ \ {\rm (fixed)}\ ,\nonumber \\
m_{s} &=& 0.598 GeV\ ,\nonumber \\
m_{c}&=&1.406 GeV\ ,\nonumber \\
m_{b} &=& 4.741 GeV\ ,
\label{fit_npvc}\\
\alpha_{s} &=& 0.539\ ,\nonumber\\
b&=&0.142 GeV^{2}\ \ \ {\rm (fixed)}\ ,\nonumber
\end{eqnarray}
to obtain an excellent fit to the spin averaged heavy-light states,
as shown in Table \ref{tall}.

\subsection{Relativistic flux tube confinement (RFTC)}

In the RFTC model
formalism for fermionic
quark confinement  is unusual in that the confinement
is introduced into the kinetic  rather than in the usual interaction
term. The flux tube contributes to both
energy and momentum, so it makes little sense to consider
it as a ``potential'' type interaction. By a covariant
substitution we add the tube to the quark momentum and energy.
We may equivalently view this as a ``minimal
substitution'' of a vector interaction field. The result
nicely reduces to the Nambu string in the limit in which
the quark moves ultra-relativistically.
This physically motivated generalization of the potential model
incorporates many aspects of QCD \cite{fermionic}.

In this model the $I\hspace{-1.65mm}L$ matrix is the same as in (\ref{llnpvc}),
while
the interaction matrix $I\hspace{-1.65mm}I$ is given by
\begin{equation}
I\hspace{-1.65mm}I =  \left(
\begin{array}{cc}
-\frac{4\alpha_{s}}{3r}+H_{t}^{k} & T_{t} \\
T_{t} & -\frac{4\alpha_{s}}{3r}+H_{t}^{-k}
\end{array}
\right)\  .
\end{equation}
In the above $T_{t}$ is defined as
\begin{equation}
T_{t} = \frac{1}{2} \left[ \frac{(1-k)}{\sqrt{-k(1-k)}} p^{-k}_t
-\frac{(1+k)}{\sqrt{k(1+k)}} p^k_t \right] \ .
\end{equation}
The flux tube energy and momentum, obtained by symmetrization of
the classical expressions \cite{lacourse, aft}, are defined by
($\{A,B\}=AB+BA$)
\begin{eqnarray}
H_{t}^{\pm k}&=&\frac{a}{2}
\{r,\frac{\arcsin{v_{\perp}^{\pm k}}}{v_{\perp}^{\pm k}}\}\ ,
\label{htube}\\
p_{t}^{\pm k}&=& = a \{r ,F(v_{\perp}^{\pm k})\}\  ,\label{ptube}
\end{eqnarray}
with ($\gamma_{\perp} = 1/\sqrt{1-v_{\perp}^{2}}$)
\begin{equation}
F(v_{\perp})=\frac{1}{4v_{\perp}}(\frac{\arcsin{v_{\perp}}}{v_{\perp}}
-\frac{1}{\gamma_{\perp}}) \ .
\end{equation}
The only unknown operators in the above expressions
are $v_{\perp}^{\pm k}$.
These are determined from the heavy-light orbital angular momentum
equation as in the spinless RFT model \cite{aft}. With the
definition $W_{r}=\sqrt{p_{r}^{2}+m^{2}}$, these equations are
\cite{fermionic}
\begin{eqnarray}
&\left[  \frac{\sqrt{k(k+1)}}{r}  =
\frac{1}{2}\{W_{r},\gamma_{\perp}^{k}v_{\perp}^{k}\}+
a \{r,F(v_{\perp}^{k})\}\ \right]
f_{j}^{k}(r){\cal Y }_{jm}^{k}(\hat{{\bf r}}) \  ,& \label{eq:angmom1} \\
&\left[ \frac{\sqrt{-k(-k+1)}}{r}  =
\frac{1}{2}\{W_{r},\gamma_{\perp}^{-k}v_{\perp}^{-k}\}+
a \{r,F(v_{\perp}^{-k})\}\ \right]
g_{j}^{k}(r){\cal Y }_{jm}^{-k}(\hat{{\bf r}})\ .& \label{eq:angmom2}
\end{eqnarray}
The numerical technique used to solve for $v_{\perp}$ is discussed in detail
elsewhere \cite{aft}.

Theoretical predictions of the model with
parameters
\begin{eqnarray}
m_{u,d}&=&0.300 \ GeV \ {\rm (fixed)}\ ,\nonumber \\
m_{s}&=&0.580\ GeV\ ,\nonumber \\
m_{c}&=&1.350\ GeV\ ,\nonumber \\
m_{b}&=&4.685\ GeV\ ,\label{fit_rftc}\\
b&=&0.181\ GeV^{2}\ {\rm (fixed)}\ ,\nonumber \\
\alpha_{s}&=&0.508\ , \nonumber
\end{eqnarray}
are shown in Table \ref{tall}.
Again, the agreement with spin-averaged experimental masses is very good.

We conclude this section by noting that all of the above models
have been used for the calculation of the elastic IW form factor
\cite{observations,fermionic},
 and the predicted IW functions ($\xi_{C}(\omega)$) were all quite
consistent with the experimental data \cite{argus,cleo2}.

\section{Branching ratios and comparison with other results}
\label{res}

Our results for branching ratios obtained from the three different models
discussed in Section \ref{hl} are shown in Table \ref{tab1}.
We have assumed  here that
$V_{cb}= 0.040$ and $\tau_{B} = 1.5ps$. Table
\ref{tab2} contains a comparison of our DESC calculation with calculation
of Scora et al. (ISGW2 model) \cite{isgw2}, the one of Suzuki et al.
(SISM model) \cite{suzuki}, with the QCD sum rule approach
of Colangelo et al. (CNP model) \cite{cnp}, and with the
model of Sutherland et al. (SHJL model) \cite{holdom}. It is
worth noting that results quoted for SISM  and CNP are also
obtained in the heavy quark limit.  Table \ref{tab3}
contains ratios of partial widths for $B$ decays into members
of the same $D^{**}$ doublet.

For calculation of branching ratios we have used experimental
meson masses wherever possible. In those cases the only model dependent
inputs were the appropriate IW form factors. For  decays where the
$D^{**}$ mass is unknown, we have used spin-averaged masses obtained
in a specific model. Based on the available information on the
splitting between $D$ and $D^{*}$ (or $D_{1}$ and $D_{2}^{*}$), one
could estimate mass splitting in other excited doublets, and use that
together with model dependent spin-averaged mass  to obtain
separate prediction for the mass of each member of that doublet.
Meson masses obtained in this way
could then be used in the calculation of the
 branching ratio for the corresponding
decay.
However, we have found that this procedure does not significantly
affect  the results. For example, using
spin-averaged mass of $1974\ MeV$ for
$D(1867)$ and $D^{*}(2009)$, instead of their experimental masses,
in the case of  DESC yields branching ratios of 2.235\% and 6.773\%
instead of 2.401\% and 6.615\%, which are given in Table \ref{tab1}.
For the higher states this effect is even less noticeable.

Let us first discuss $B\rightarrow D$ and $B\rightarrow D^{*}$ transitions.
Recent results from CLEO \cite{cleo3},
\begin{eqnarray}
BR(B^{-}\rightarrow D^{0}e^{-}\bar{\nu}_{e}) &=& (1.95\pm 0.55)\%\
,\label{bd1}\\
BR(B^{-}\rightarrow D^{*0}e^{-}\bar{\nu}_{e}) &=& (5.13\pm 0.84)\%\
,\label{bd2}
\end{eqnarray}
and results given in Tables \ref{tab1} and \ref{tab2} imply that
all models we used, as well as the ISGW2 and SHJL models,
require $V_{cb}$ slightly
lower than $0.040$. In our models values range from about
0.036 for DESC, to about 0.038 for RFTC and SEVC. On the other hand,
ISGW2 is consistent with 0.035, SHJL gives about 0.036, and
SISM and CNP models agree with $V_{cb}$ of about 0.040.
{}From Table \ref{tab2} it can also be found that\footnote{It is worth noting
that the calculation of  \cite{suzuki} used
form factor definitions which are not consistent with the
covariant trace formalism.}
\begin{equation}
\frac{BR(B^{-}\rightarrow D^{0} e^{-}\bar{\nu}_{e})}
{BR(B^{-}\rightarrow D^{*0} e^{-}\bar{\nu}_{e})} =
\left\{ \begin{array}{ll}
0.48 \ ,& {\rm for\ ISGW2}\\
0.34 \ ,& {\rm for \ SISM} \\
0.33 \ ,& {\rm for \ CNP}\\
0.31\ , & {\rm for \ SHJL}\end{array}
\right.
\ .
\end{equation}
On the other hand, our calculation with three different models yields
(see Table \ref{tab3})
\begin{equation}
\frac{BR(B^{-}\rightarrow D^{0} e^{-}\bar{\nu}_{e})}
{BR(B^{-}\rightarrow D^{*0} e^{-}\bar{\nu}_{e})} =
\left\{ \begin{array}{ll}
0.35 \ ,& {\rm for\ RFTC}\\
0.36 \ ,& {\rm for \ DESC} \\
0.35 \ ,& {\rm for \ SEVC}\end{array}
\right.
\ .
\end{equation}
The results quoted in
 (\ref{bd1}) and (\ref{bd2})
imply an experimental ratio of
\begin{equation}
\frac{BR(B^{-}\rightarrow D^{0} e^{-}\bar{\nu}_{e})}
{BR(B^{-}\rightarrow D^{*0} e^{-}\bar{\nu}_{e})} = 0.38 \pm 0.17\ .
\end{equation}
 It is interesting
to note that ratio of polarization states of
$D$ and $D^{*}$ is $ 0.33$.

Individual contributions of $P$-wave $j=\frac{3}{2}$ states to the total
semi-leptonic decay rate is
another interesting point.
{}From Table \ref{tab1}  it can be seen that  the
total semi-leptonic branching ratio for
 $B\rightarrow D_{1}$ and $B\rightarrow D_{2}^{*}$
is expected to be
\begin{equation}
BR(B^{-}\rightarrow D_{1}^{0} e^{-}\bar{\nu}_{e}) +
BR(B^{-}\rightarrow D_{2}^{*0} e^{-}\bar{\nu}_{e}) =
\left\{ \begin{array}{ll}
0.84 \left|\frac{V_{cb}}{0.040}\right|^{2}\frac{\tau_{B}^{}}{1.50ps}\%
\ ,& {\rm for\ RFTC}\\
0.69 \left|\frac{V_{cb}}{0.040}\right|^{2}\frac{\tau_{B}^{}}{1.50ps}\%
\ ,& {\rm for \ DESC} \\
0.79 \left|\frac{V_{cb}}{0.040}\right|^{2}\frac{\tau_{B}^{}}{1.50ps}\%
\ ,& {\rm for \ SEVC}\end{array}
\right.
\ .
\end{equation}
These results are slightly larger than the ISGW2 result of
 $0.65
\left|\frac{V_{cb}}{0.040}\right|^{2}\frac{\tau_{B}^{}}{1.50ps}\%$,
and significantly disagree with SISM, CNP, and SHJL models
($0.20$, $0.37$, and $0.46\
\left|\frac{V_{cb}}{0.040}\right|^{2}\frac{\tau_{B}^{}}{1.50ps}\%$,
respectively). However, as one can see from Table \ref{tab3},
our ratios of these two $P$ wave decays are qualitatively different
from the one obtained in ISGW2 and SHJL, and agree with SISM and
CNP models. We find
\begin{equation}
\frac{BR(B^{-}\rightarrow D_{1}^{0} e^{-}\bar{\nu}_{e})}
{BR(B^{-}\rightarrow D_{2}^{*0} e^{-}\bar{\nu}_{e})} =
\left\{ \begin{array}{ll}
0.60 \ ,& {\rm for\ RFTC}\\
0.63 \ ,& {\rm for \ DESC} \\
0.60 \ ,& {\rm for \ SEVC}\end{array}
\right.
\ ,
\end{equation}
while the other models yield
\begin{equation}
\frac{BR(B^{-}\rightarrow D_{1}^{0} e^{-}\bar{\nu}_{e})}
{BR(B^{-}\rightarrow D_{2}^{*0} e^{-}\bar{\nu}_{e})} =
\left\{ \begin{array}{ll}
2.00 \ ,& {\rm for\ ISGW2}\\
0.70 \ ,& {\rm for \ SISM} \\
0.50 \ ,& {\rm for \ CNP} \\
1.54 \ ,&{\rm for \ SHJL} \end{array}
\right.
\ .
\end{equation}
Again, it is interesting to note that
the ratio of number of polarization states of
$D_{1}$ and $D_{2}^{*}$ is 0.6.
It remains to be seen whether
this discrepancy between our results (which are obtained in the heavy-light
limit) and ISGW2 and SHJL\footnote{In \cite{holdom} one can also
find results obtained in the heavy quark limit. These are in general
much smaller than our results, but the ratios of partial widths
for the $B$ decays into the members of the same $D^{**}$ doublet
agree much better with our predictions.} models can be explained with
large $\frac{1}{m_{c}}$
effects \cite{isgurp}.

As already mentioned, within the HQET framework
the only model dependent input for the decays
$C\rightarrow C,C^{*}$ and $C\rightarrow F, F^{*}$ are the unknown IW
functions.
For these decays the uncertainty introduced by using a particular model is
about 10\%.
For other decays, $D^{**}$ mass is not known, so that in the calculation
of  decay rates we used predictions of a particular model. Therefore,
one should expect larger discrepancies between different models.
{}From Table \ref{tab1} one can see that this is indeed the case.
 However, results
obtained from the three different models are
not significantly different. For example, for the decays
$C\rightarrow E,E^{*}$ uncertainty the introduced by using a specific model
is about 20\%. Also, as one can see from Table \ref{tab3}, the ratios
of the two exclusive decay widths for  members of the same doublet
are all consistent, which is the consequence of the application
of HQET.

\section{Fractional semi-leptonic decay rates}
\label{res2}

The exclusive decay rates discussed earlier suffer from a variety
of theoretical oversimplifications. Some of the things which were not taken
into account are QCD corrections, spectator effects, and deviations
from  exact heavy quark symmetry. In addition, there are
several parameters
which need to be specified before definite predictions can be made.
Among these are the CKM parameter $V_{cb}$, the $b$-quark lifetime,
and the quark masses.

Many of the above problems can be reduced by considering
fractions of the inclusive
$b\rightarrow c e \bar{\nu}_{e} $ rate. In particular, $V_{cb}$ exactly
cancels. Also, since
 the sum of the exclusive
rates  equals the inclusive rate,
and since the inclusive calculation is structurally similar to the
exclusive ones, there should be  some cancellation of the
QCD,  spectator corrections, and heavy quark mass dependence.
Since the inclusive rate has been measured, one can directly
compare these fractional predictions with experiment in several cases.

The inclusive spectator model decay rate
for $b\rightarrow c e \bar{\nu}_{e}$ is \cite{altarelli, bigi}
\begin{equation}
\Gamma(b\rightarrow c e \bar{\nu}_{e}) = \frac{G_{F}^{2} m_{b}^{5}
|V_{cb}|^{2}}
{192 \pi^{3}} I(\frac{m_{c}^{2}}{m_{b}^{2}},0,0)\ ,
\end{equation}
where \cite{cortes}
\begin{equation}
I(x,0,0)= (1-x^{2})(1-8 x + x^{2}) - 12 x^{2}\ln{x}\ .
\end{equation}
If for the moment we ignore the oversimplifications of the above
inclusive model and assume $V_{cb}=0.040$, the $b$-quark lifetime
of $\tau_{b}=1.5\times 10^{-12} s$, and the quark masses of the three
realistic models given in (\ref{fit_desc}), (\ref{fit_npvc}), and
(\ref{fit_rftc}), we find a total branching ratio
\begin{equation}
BR(b\rightarrow c e \bar{\nu}_{e}) =
\left\{\begin{array}{ll}
10.27 \left|V_{cb}/0.040\right|^{2}(\tau_{B}/1.50ps)\%
 \ ,& {\rm for\ RFTC}\\
10.31  \left|V_{cb}/0.040\right|^{2}(\tau_{B}/1.50ps)\%
\ ,& {\rm for \ DESC} \\
10.53  \left|V_{cb}/0.040\right|^{2}(\tau_{B}/1.50ps)\%
\ ,& {\rm for \ SEVC}\end{array}
\label{brfree}
\right.\ .
\end{equation}
The experimental branching ratio,
\cite{gronberg}
\begin{equation}
BR(B^{-}\rightarrow Xe^{-}\bar{\nu}_{e}) = (10.49\pm 0.46)\%\ ,
\label{brexp}
\end{equation}
is in excellent agreement with the above numbers. However,
one should keep in mind that the predicted value is
very sensitive to the choice of $V_{cb}$, $\tau_{b}$, and quark masses.

One plausibly assumes that ratio of the exclusive branching ratios to the
inclusive one,
\begin{equation}
R^{**} = \frac{BR(B\rightarrow D^{**}e \bar{\nu}_{e})}{
BR(b\rightarrow ce \bar{\nu}_{e})}\ ,
\end{equation}
will be more accurate than either of these separately. We first apply this
idea to $B\rightarrow D$ and $B\rightarrow D^{*}$ decays. From (\ref{bd1}),
(\ref{bd2}) and (\ref{brexp}) we find experimental fractions,
\begin{eqnarray}
R_{D}^{exp}& =& 0.19\pm 0.06\ ,
\label{r1}\\
R_{D^{*}}^{exp} &=& 0.49\pm 0.10
\label{r2}\ .
\end{eqnarray}
{}From Table \ref{tab1} one can see that our three models predict
\begin{equation}
R_{D}^{th} =
\left\{\begin{array}{ll}
0.20 \ ,& {\rm for\ RFTC}\\
0.23  \ ,& {\rm for \ DESC} \\
0.20  \ ,& {\rm for \ SEVC}\end{array}
\right.\ ,
\end{equation}
and
\begin{equation}
R_{D^{*}}^{th} =
\left\{\begin{array}{ll}
0.58 \ ,& {\rm for\ RFTC}\\
0.64  \ ,& {\rm for \ DESC} \\
0.58 \ ,& {\rm for \ SEVC}\end{array}
\right.\ .
\end{equation}
The predicted values are reasonably consistent with measurement in all cases.

The fraction of semi-leptonic decay into final states other than
$D$ or $D^{*}$ is by (\ref{r1}) and (\ref{r2})
\begin{eqnarray}
R^{exp}(D^{**}\ {\rm other\ than\ }D{\rm \ and \ }D^{*}) &=&
1-(R_{D}^{exp}+R_{D^{*}}^{exp})
\nonumber \\
& =& 0.32\pm 0.16\ .
\end{eqnarray}
{}From Table \ref{tab1} we see that three models discussed in this paper imply
\begin{eqnarray}
R^{th}(D^{**}\ {\rm other\ than\ }D{\rm \ and \ }D^{*})& =&
1-(R_{D}^{th}+R_{D^{*}}^{th})\nonumber \\
&=&
\left\{\begin{array}{ll}
0.22 \ ,& {\rm for\ RFTC}\\
0.13  \ ,& {\rm for \ DESC} \\
0.22 \ ,& {\rm for \ SEVC}\end{array}
\right.\ .
\end{eqnarray}

It is interesting to observe that single excited charmed states alone are
nearly
consistent with accounting for  the entire inclusive semi-leptonic
decay fraction. As a more direct way of seeing this note
the total fractional percentage in Table \ref{tab2}. The predicted fraction
into all $D^{**}$ states lies at $89\%$ or above for the three models
discussed in this paper.

\section{Consistency with  sum rules}
\label{bj}

\subsection{The Bjorken sum rule}

The Bjorken sum rule \cite{bjsum,bjsum2} relates the derivative of the
elastic form factor to the values of inelastic
$S$- to $P$-wave form factors at the zero recoil point. In our notation,
\begin{equation}
-\xi'_{C}(1) = \frac{1}{4} +
\frac{1}{4}\sum_{i}\left| \xi_{E}^{(i)}(1)\right|^{2}
+\frac{2}{3}\sum_{j}\left| \xi_{F}^{(j)}(1)\right|^{2}\ .
\label{bjsr}
\end{equation}
Since the $S$- to $P$-wave form factor normalizations
at the zero recoil point are not fixed (as is the case of the elastic
form factor $\xi_{C}$), but instead depend on the energies and wave functions
of the LDF \cite{modelling,stop},
it is not immediately obvious that form factors obtained from
the three different models will also satisfy the Bjorken sum rule.
In particular, we observe from (\ref{bjsr}) the manifestly valid
inequality
\begin{equation}
-\xi_{C}'(1)\geq \frac{1}{4}\ .
\end{equation}
On the other hand, it follows from (\ref{xic}) that
\cite{modelling,sadzikowski}
\begin{equation}
-\xi_{C}'(1)\geq \frac{1}{2}\ .
\end{equation}
Combined together, these two bounds imply
\begin{equation}
\frac{1}{4}\sum_{i}\left| \xi_{E}^{(i)}(1)\right|^{2}
+\frac{2}{3}\sum_{j}\left| \xi_{F}^{(j)}(1)\right|^{2}
\geq \frac{1}{4}\ .
\label{psum}
\end{equation}
One may ask here whether the above two bounds for
$-\xi'_{C}(1)$ are consistent, or can
one devise a model for which the $S$- to $P$-wave form factors in
(\ref{psum}) come to less than $\frac{1}{4}$ \cite{modelling}.

With this in mind we  have evaluated numerically both sides of
(\ref{bjsr}) in order to check self consistency of the three
models used in this paper. In the sum of the right-hand side of
(\ref{bjsr}) we included the lowest $P$-waves
plus the first two radial excitations.
For the Dirac equation with scalar confinement, with parameters given in
(\ref{fit_desc}), the
direct evaluation yields
\begin{equation}
-\xi'_{C}(1) \simeq 0.86\ ,
\end{equation}
while the sum rule approach (the right-hand side of (\ref{bjsr})) gives
\begin{equation}
-\xi'_{C}(1)  =0.25 +(0.535 + 0.026 + 0.004 +\ldots) \simeq 0.82\ .
\end{equation}
For the
Salpeter equation with vector confinement, with parameters given in
(\ref{fit_npvc}), we obtained after
direct evaluation
\begin{equation}
-\xi'_{C}(1) \simeq 1.26\ ,
\end{equation}
and the sum rule result was
\begin{equation}
-\xi'_{C}(1)  \simeq 1.03\ .
\end{equation}
Similarly, the relativistic flux tube model confinement,
 with parameters given in
(\ref{fit_rftc}), yields after the
direct evaluation
\begin{equation}
-\xi'_{C}(1) \simeq 1.14\ ,
\end{equation}
while the sum rule approach gives
\begin{equation}
-\xi'_{C}(1)  \simeq 1.09\ .
\end{equation}
It is also worthwhile noting here the CLEO result for derivative
of the  renormalized
form factor \cite{cleo2}
\begin{equation}
-\hat{\xi}'_{C}(1) = 0.84\pm 0.12\pm 0.08\ .
\end{equation}

Since all the terms on the right-hand side of (\ref{bjsr})
are positive definite, and since we are neglecting nonresonant
contributions to final states containing a pion (even if those
were small), one might argue that in a self-consistent model
result for $-\xi'_{C}(1)$
obtained by direct calculation should be smaller than the
one obtained in the sum rule approach.
Indeed, this is what happens in all three of the heavy-light
models used in this paper. Furthermore, for all three models
(and especially for the DESC and RFTC) the sum rule is very close
to being saturated by the resonant contributions.

However, we must also mention at this point that a similar
calculation was also performed
in \cite{holdom}. These authors find (also in the heavy quark limit)
\begin{equation}
-\xi'_{C}(1)  = 1.28\ ,
\end{equation}
after the direct calculation,
and
\begin{equation}
-\xi'_{C}(1) =0.25 + 0.21+\ldots \simeq 0.46\ ,
\end{equation}
in the sum rule approach (using only the lowest $P$-wave mesons).
The above result shows that the sum rule is far from being
saturated by resonances. The difference between the two approaches
was in \cite{holdom} explained
as being mainly due  to nonresonant contributions to final states
containing a pion.

\subsection{The Voloshin sum rule}

The Voloshin sum rule \cite{voloshin} is the analog of the ``optical''
sum rule for the dipole scattering of light in atomic physics.
In terms of our form factors and energies of the LDF
($E_{D^{**}}= m_{D^{**}}-m_{c}$), it can be written in the form
\begin{equation}
 \frac{1}{2}
=
\frac{1}{4}\sum_{i}(\frac{E_{D^{**}}^{(i)}}{E_{D}}-1)
\left| \xi_{E}^{(i)}(1)\right|^{2}
+\frac{2}{3}\sum_{j}(\frac{E_{D^{**}}^{(j)}}{E_{D}}-1)
\left| \xi_{F}^{(j)}(1)\right|^{2}\equiv \Delta\ .
\label{vosr}
\end{equation}
Here $E_{D^{**}}^{(i)}$ and $E_{D^{**}}^{(j)}$ denote
energies of the LDF of the $j=\frac{1}{2}$ and $j=\frac{3}{2}$ $P$-wave
mesons, respectively. In this sum rule
the dependence on the unknown charm mass (or the LDF energy)
is even stronger  than in the case of the Bjorken sum rule,
where it is contained only implicitly through  form factors.
Therefore, one could expect that model calculations used with (\ref{vosr})
are less satisfactory than in the case of (\ref{bjsr}).
Still, one could again make the same arguments as in the case
of the Bjorken sum rule, and conclude that any self-consistent
model calculation of the right-hand side of (\ref{vosr}) should
yield result smaller than $0.5$.

To evaluate the right-hand side of (\ref{vosr}),
we have used the spin-averaged energies of the LDF in $D$ and $D^{*}$ and
other $D^{**}$ doublets
(because the spin-averaged experimental masses were used in numerical
calculations which determined
quark masses). As before,   we  have used
the lowest $P$-waves
plus the first two radial excitations for both, $j=\frac{1}{2}$ and
$j=\frac{3}{2}$ mesons.

For the Dirac equation with scalar confinement (DESC),
 with parameters given in
(\ref{fit_desc}), we find
\begin{equation}
\Delta = 0.395 + 0.033+0.007+\ldots \simeq 0.44\ ,
\end{equation}
which is smaller than the predicted value of 0.5.
On the other hand, the Salpeter equation with vector confinement (SEVC),
 with parameters given in
(\ref{fit_npvc}),   results in
\begin{equation}
\Delta \simeq 0.63\ ,
\end{equation}
while the relativistic flux tube model confinement (RFTC),
 with parameters given in
(\ref{fit_rftc}), yields
\begin{equation}
\Delta \simeq 0.60 \ .
\end{equation}
These two results are about 20\% higher than the predicted value of 0.5.
Again, we emphasize that in the case of the Voloshin sum rule
numerical results
are much more dependent
on the unknown charm quark mass and other parameters of the model.
Unfortunately, we are not aware of any other model
which has used the Voloshin sum rule as a test of
self-consistency, so that we cannot compare our result with the
literature.
Nevertheless, we do have to say that two out of three  heavy-light models
used here (SEVC and RFTC)
appear to be inconsistent as far as Voloshin sum rule is concerned.
Based on  the above calculations,
one might also argue that the DESC
model predictions are the most reliable of all the
results presented in this paper.

\section{Conclusion}
\label{con}

We have examined the role of semi-leptonic $B$ decay into higher
charmed mesons. Within a HQET framework we have
evaluated branching ratios for $B\rightarrow D^{**}e\bar{\nu}_{e}$, where
the $D^{**}$ are all $S$- and $P-$wave mesons,  $D-$wave
mesons with $j=\frac{3}{2}$, and some of their radial excitations.
Our numerical calculations
are based upon three realistic models. In each case
a light fermion interacts with a fixed source. A short distance
vector Coulomb interaction is used, and at large distances
the fermion is confined by a scalar, time-component vector,
or a flux tube.\footnote{Results based on spinless models
such as semi-relativistic quark model (also used in \cite{stop}),
or spinless relativistic flux tube model (used in \cite{rftiwf})
are not significantly different from the results based on
the models discussed in this paper.}

An important aim here has been to determine how much of the
semi-leptonic decay rate ends up as one of the higher charmed resonances.
Considered as a fraction of the inclusive $b\rightarrow c e \bar{\nu}_{e}$
rate we find that between 90 and 95\% of $B$ mesons decay
semi-leptonically into a single excited charmed resonance.

Another area which we have
explored  in this paper is the consistency of form factors obtained
from three different models with the Bjorken and Voloshin
sum rules. One might
suspect that there may be an inconsistency between the Bjorken expression
for the slope of the Isgur-Wise function at the zero recoil point, and
the expression for the same quantity which is obtained in
the wave function approach and depends only on the elastic
($S$- to $S$-wave) transition. The two methods
yield different
upper bounds for the IW slope. However, using
numerical values for
the lowest two $S$- to $P$-wave form factors
(and also for the derivative of elastic form factor)
 at the zero recoil point,
 we find
that all three models yield results consistent with the Bjorken sum rule.
The Voloshin sum rule provides another sensitive test of model
 calculations. There, the dependence on the unknown charm
quark mass is even stronger than in the case of the Bjorken sum rule, and
one might expect that $\frac{1}{m_{c}}$ effects play an even more
significant role. All three of the heavy-light models used in this paper
give results which are slightly higher than the sum rule prediction.
Nevertheless, given the large qualitative differences
between models itself, and the fact that their predictions
are very similar in all cases, we believe that all the
results presented in this paper are trustworthy.

We have also compared our results with other calculations
available in the literature. We find significant
disagreements with \cite{isgw2} and \cite{holdom} in ratios of
decay widths for
$B$ decays into the members of the same $D^{**}$ doublet. On the other hand,
our results are in general
significantly larger than the ones obtained in \cite{suzuki} and
\cite{cnp}.

\vskip 1cm
\begin{center}
ACKNOWLEDGMENTS
\end{center}
We would like to thank B. Holdom for helpful comments.
This work was supported in part by the U.S. Department of Energy
under Contract No. DE-FG02-95ER40896 and in part by the University
of Wisconsin Research Committee with funds granted by the Wisconsin Alumni
Research Foundation.

\newpage

\begin{table}
\vspace*{-3cm}
\normalsize
\begin{center}
TABLES
\end{center}
\caption{ Heavy-light states.
Spin-averaged experimental
masses are calculated in the usual way, by taking $\frac{3}{4}$
($\frac{5}{8}$) of
the triplet and $\frac{1}{4}$ ($\frac{3}{8}$) of the singlet mass for the
$S(P)$  waves).
Theoretical
results
 are obtained from the three heavy-light models
discussed in the text. Model parameters are given
in (\protect\ref{fit_desc}) (DESC), (\protect\ref{fit_npvc}) (SEVC), and
(\protect\ref{fit_rftc}) (RFTC).
Theoretical errors with respect to the spin-averaged
experimental masses are shown in parentheses. }
\vskip 0.2cm
\begin{center}
\begin{tabular}{|lllrl|l|l|l|}
\hline
\hline
         State
       & \multicolumn{3}{c}{}
       & Mass
       & RFTC
       & DESC
       & SEVC
\\
       & \hspace*{-5pt}$J^{P}_{j}$
       & \hspace*{-5pt}$k$
       & \hspace*{-5pt}$^{2S+1}L_{J}$
       & (MeV)
       & (MeV)
       & (MeV)
       & (MeV)
\\
\hline
         \underline{$c\bar{u},\ c\bar{d}$ quarks}
       &
       &
       &
       &
       &
       &
       &
\\
         $\begin{array}{ll}
              		D       (1867) &C\\
   			D^{*}   (2009) &C^{*}\end{array}$
       & \hspace*{-5pt}$\begin{array}{l}
       			0^{-}_{\frac{1}{2}} \\
      			1^{-}_{\frac{1}{2}} \end{array}$
       & \hspace*{-5pt}$ -1$
       &
	 \hspace*{-5pt}$\left. \begin{array}{l}
			\hspace{+1.1mm}   ^{1}S_{0} \\
			\hspace{+1.1mm}   ^{3}S_{1} \end{array}\right] $
       & $1S\ (1974)$
       & $1981(+7)$
       & $1977(+3)$
       & $1980(+6)$
\\
	 $\begin{array}{ll}
			D_{1}     (2425) &F\\
			D_{2}^{*}  (2459) &F^{*}\end{array}$
       & \hspace*{-5pt}$\begin{array}{l}
			1^{+}_{\frac{3}{2}} \\
			2^{+}_{\frac{3}{2}} \end{array}$
       & \hspace*{-5pt}$-2$
       & \hspace*{-5pt}$\left.\begin{array}{r}
			\hspace{+0.5mm}^{1}P_{1}/^{3}P_{1} \\
			\hspace{+0.5mm}^{3}P_{2} \end{array}\right] $
       & $1P\ (2446)$
       & $2439(-7)$
       & $2444(-2)$
       & $2440(-6)$
\\
         \underline{$c\bar{s}$ quarks}
       &
       &
       &
       &
       &
       &
       &
\\
	 $\begin{array}{ll}
   			D_{s}  (1969) &C \\
   			D^{*}_{s}  (2112) &C^{*}\end{array}$
       & \hspace*{-5pt}$\begin{array}{l}
      			0^{-}_{\frac{1}{2}} \\
      			1^{-}_{\frac{1}{2}} \end{array}$
       & \hspace*{-5pt}$-1$
       & \hspace*{-5pt}$\left. \begin{array}{l}
			\hspace{+1.1mm}    ^{1}S_{0} \\
			\hspace{+1.1mm}    ^{3}S_{1} \end{array}\right] $
       & $1S\ (2076)$
       & $2071(-5)$
       & $2074(-2)$
       & $2072(-4)$
\\
	 $\begin{array}{ll}
			D_{s1}  (2535) &\hspace*{-1.5mm}F \\
		        D_{s2}^{*}  (2573) &\hspace*{-1.5mm}F^{*}\end{array}$
       & \hspace*{-5pt}$\begin{array}{l}
			1^{+}_{\frac{3}{2}} \\
			2^{+}_{\frac{3}{2}} \end{array}$
       & \hspace*{-5pt}$-2$
       & \hspace*{-5pt}$\left.\begin{array}{r}
			\hspace{+0.5mm}^{1}P_{1}/^{3}P_{1} \\
		  	\hspace{+0.5mm}^{3}P_{2} \end{array}\right] $
       & $1P\ (2559)$
       & $2564(+5)$
       & $2560(+1)$
       & $2564(+5)$
\\
         \underline{$b\bar{u},\ b\bar{d}$ quarks}
       &
       &
       &
       &
       &
       &
       &
\\
         $\begin{array}{ll}
   			B      (5279) &C \\
   			B^{*}   (5325) &C^{*}\end{array}$
       & \hspace*{-5pt}$\begin{array}{l}
     	 		0^{-}_{\frac{1}{2}} \\
      			1^{-}_{\frac{1}{2}} \end{array}$
       & \hspace*{-5pt}$-1$
       & \hspace*{-5pt}$\left. \begin{array}{l}
			\hspace{+1.1mm}   ^{1}S_{0} \\
 			\hspace{+1.1mm}   ^{3}S_{1} \end{array}\right] $
       & $1S\ (5314)$
       & $5316(+2)$
       & $5313(-1)$
       & $5316(+2)$
\\
         \underline{$b\bar{s}$ quarks}
       &
       &
       &
       &
       &
       &
       &
\\
         $\begin{array}{ll}
			B_{s}      (5374) &C\\
			B_{s}^{*}  (5421) &C^{*}\end{array}$
       & \hspace*{-5pt}$\begin{array}{l}
			0^{-}_{\frac{1}{2}} \\
			1^{-}_{\frac{1}{2}} \end{array} $
       & \hspace*{-5pt}$-1$
       & \hspace*{-5pt}$\left.\begin{array}{l}
			\hspace{+1.0mm} ^{1}S_{0} \\
			\hspace{+1.0mm} ^{3}S_{1} \end{array}\right] $
       & $1S\ (5409)$
       & $5407(-2)$
       & $5410(+1)$
       & $5407(-2)$
\\
\hline
\hline
\end{tabular}
\end{center}
\label{tall}
\end{table}

\begin{table}
\caption{Exclusive partial widths for decays
$B\rightarrow D^{**}e\bar{\nu}_{e}$ obtained from the three
different models discussed in the paper.
$\Gamma$ is given in units of
$\left[\left|\frac{V_{cb}}{0.040}\right|^{2}10^{-15}GeV\right]$,
$BR$ is in units of $\left[
\left|\frac{V_{cb}}{0.040}\right|^{2}\frac{\tau_{B}^{}}{1.50ps}\%
\right]$, while the ratio $R= \frac{BR(B\rightarrow D^{**}e\bar{\nu}_{e})}{
BR(b\rightarrow ce\bar{\nu}_{e})}$ is given in $[\%]$.
Numerical values of $BR(b\rightarrow ce\bar{\nu}_{e})$
for a particular model can be found
in (\protect\ref{brfree}). }
\label{tab1}
\begin{center}
\begin{tabular}{|cc|ccc|ccc|ccc|}
\hline
\hline
\multicolumn{2}{|c|}{State}&
\multicolumn{3}{c|}{RFTC} &
\multicolumn{3}{c|}{DESC} &
\multicolumn{3}{c|}{SEVC} \\
$D^{**}$    &   $J^{P}_{j}$  &
$\Gamma $& $BR$ & $R$ &
$\Gamma $& $BR$ & $R$ &
$\Gamma $& $BR$ & $R$ \\
\hline
 $C$          &$0^{-}_{\frac{1}{2}}$
& 9.026 & 2.057 & 20.03
&10.54 &2.401  & 23.29
&9.420 &2.147  & 20.40   \\
$C^{*}$      &$1^{-}_{\frac{1}{2}}$
 &26.09 & 5.946 & 57.89
&29.03 & 6.615 & 64.14
&26.80 & 6.108  & 58.03  \\
$E$         &$0^{+}_{\frac{1}{2}}$
& 0.211 & 0.048 & 0.468
& 0.303 & 0.069 & 0.670
& 0.204 & 0.047 & 0.442 \\
$E^{*}$      &$1^{+}_{\frac{1}{2}}$
& 0.299 & 0.068 & 0.663
&0.419 & 0.096  & 0.926
&0.289 & 0.066   & 0.626  \\
$F$         &$1^{+}_{\frac{3}{2}}$
& 1.383 & 0.315 & 3.069
&1.161&0.265  & 2.565
&1.294 & 0.295  & 2.802   \\
$F^{*}$      &$2^{+}_{\frac{3}{2}}$
&2.307 & 0.526 & 5.119
&1.854 &0.423  & 4.097
&2.152 & 0.490  & 4.659   \\
$G$         &$1^{-}_{\frac{3}{2}}$
& 0.016 & 0.004 & 0.036
&0.023 & 0.005  & 0.051
& 0.016 & 0.004 & 0.035  \\
$G^{*}$      &$2^{-}_{\frac{3}{2}}$
& 0.016 & 0.004 & 0.036
&0.023 & 0.005 & 0.051
&0.016 & 0.004 & 0.035 \\
$C_{2}$     &$0^{-}_{\frac{1}{2}}$
& 0.225 & 0.051 & 0.499
& 0.067 & 0.015 & 0.148
&0.215 & 0.049  & 0.466  \\
$C^{*}_{2}$  &$1^{-}_{\frac{1}{2}}$
 & 0.460 & 0.105 & 1.021
&0.131 & 0.030 & 0.289
&0.437 & 0.100 & 0.946  \\
 $E_{2}$         &$0^{+}_{\frac{1}{2}}$
&0.009 & 0.002 & 0.020
& 0.018 & 0.004 & 0.040
&0.022 & 0.005 & 0.048    \\
 $E^{*}_{2}$     &$1^{+}_{\frac{1}{2}}$
&0.011 & 0.003 & 0.024
&0.025 & 0.006 & 0.055
&0.028 &0.006  & 0.061 \\
 $F_{2}$        &$1^{+}_{\frac{3}{2}}$
& 0.068 & 0.016 & 0.151
&0.029 & 0.007 & 0.064
&0.085 & 0.019 & 0.184   \\
$F^{*}_{2}$     &$2^{+}_{\frac{3}{2}}$
& 0.101 & 0.023 & 0.224
&0.045 & 0.010 & 0.099
&0.128 &0.029  & 0.277   \\
\hline
\multicolumn{2}{|c|}{total}
&40.22 & 9.17 & 89.24
&43.67 & 9.95 & 96.49
&41.11 &9.37 & 89.01  \\
\hline
\hline
\end{tabular}
\end{center}
\end{table}

\begin{table}
\caption{Our results with the DESC model
for
$B\rightarrow D^{**}e\bar{\nu}_{e}$ compared to predictions of ISGW2
\protect\cite{isgw2}, SISM \protect\cite{suzuki}, CNP
\protect\cite{cnp}, and SHJL \protect\cite{holdom}.
$\Gamma$ is given in units of
$\left[\left|\frac{V_{cb}}{0.040}\right|^{2}10^{-15}GeV\right]$ and
$BR$ in units of $\left[
\left|\frac{V_{cb}}{0.040}\right|^{2}\frac{\tau_{B}^{}}{1.50ps}\%
\right]$. }
\label{tab2}
\begin{center}
\begin{tabular}{|cc|cc|cc|cc|cc|cc|}
\hline
\hline
\multicolumn{2}{|c|}{State}&
\multicolumn{2}{c|}{ISGW2}& \multicolumn{2}{c|}{SISM} &
\multicolumn{2}{c|}{CNP} & \multicolumn{2}{c|}{SHJL} &
\multicolumn{2}{c|}{DESC} \\
$D^{**}$    &  $J^{P}_{j}$  &
$\Gamma $& $BR$ & $\Gamma $& $BR$
& $\Gamma $& $BR$ & $\Gamma $& $BR$
& $\Gamma $& $BR$ \\
\hline
 $C$        &$0^{-}_{\frac{1}{2}}$
& 12.53 & 2.860 & 7.800 & 1.778
&7.616 & 1.736 &9.478 & 2.160
& 10.54 & 2.401 \\
$C^{*}$     &$1^{-}_{\frac{1}{2}}$
 & 26.12 &  5.950 & 23.17 & 5.279
&23.39 & 5.331 & 30.646 & 6.980
&29.03 & 6.615 \\
$E$         &$0^{+}_{\frac{1}{2}}$ &
0.316 & 0.072 & 0.118 & 0.027
&0.272 & 0.062 &0.295 & 0.067
& 0.303 & 0.069\\
$E^{*}$      &$1^{+}_{\frac{1}{2}}$ &
0.316 & 0.072 & 0.154 & 0.035
&0.381 & 0.087 & 0.421 & 0.096
& 0.419 & 0.096\\
$F$         &$1^{+}_{\frac{3}{2}}$ &
1.896 & 0.432 & 0.363 & 0.083
& 0.544 & 0.124 &1.232 & 0.281
& 1.161 & 0.265\\
$F^{*}$     &$2^{+}_{\frac{3}{2}}$ &
 0.948 & 0.216 & 0.517 & 0.118
&1.088 & 0.248 & 0.800 & 0.182
&1.854 & 0.423\\
$G$         &$1^{-}_{\frac{3}{2}}$ &
\multicolumn{2}{c|}{not given} &0.001 & 0.000
&\multicolumn{2}{c|}{not given} & 0.022 & 0.005
& 0.023 & 0.005 \\
$G^{*}$     &$2^{-}_{\frac{3}{2}}$ &
\multicolumn{2}{c|}{not given}& 0.001 & 0.000
&\multicolumn{2}{c|}{not given}& 0.003 & 0.001
& 0.023 & 0.005 \\
$C_{2}$     &$0^{-}_{\frac{1}{2}}$  &
0.000 & 0.000 & 0.071 & 0.016
&\multicolumn{2}{c|}{not given} &0.579 & 0.132
& 0.067 & 0.015\\
$C^{*}_{2}$ &$1^{-}_{\frac{1}{2}}$ &
 0.632 & 0.144 & 0.172 & 0.039
&\multicolumn{2}{c|}{not given} &\multicolumn{2}{c|}{not given}
& 0.131 & 0.030\\
 $E_{2}$         &$0^{+}_{\frac{1}{2}}$ &
\multicolumn{2}{c|}{not given}  &\multicolumn{2}{c|}{not given}
&\multicolumn{2}{c|}{not given}&\multicolumn{2}{c|}{not given}
&0.018 & 0.004\\
 $E^{*}_{2}$      &$1^{+}_{\frac{1}{2}}$ &
\multicolumn{2}{c|}{not given} &\multicolumn{2}{c|}{not given}
&\multicolumn{2}{c|}{not given}&\multicolumn{2}{c|}{not given}
&0.025 & 0.006  \\
 $F_{2}$        &$1^{+}_{\frac{3}{2}}$ &
\multicolumn{2}{c|}{not given} &\multicolumn{2}{c|}{not given}
&\multicolumn{2}{c|}{not given}&\multicolumn{2}{c|}{not given}
& 0.029 & 0.007\\
$F^{*}_{2}$      &$2^{+}_{\frac{3}{2}}$ &
\multicolumn{2}{c|}{not given} &\multicolumn{2}{c|}{not given}
&\multicolumn{2}{c|}{not given}&\multicolumn{2}{c|}{not given}
& 0.045 & 0.010\\
\hline
\multicolumn{2}{|c|}{total}
& 42.76 & 9.74 &32.36& 7.38
&33.29 & 7.59 & 43.49 &9.91
&43.67 & 9.95\\
\hline
\hline
\end{tabular}
\end{center}
\end{table}

\begin{table}
\caption{Ratios of partial widths for the $B$ decays into
the members of the same $D^{**}$ doublet obtained from the three
different models discussed in this paper (second to fourth columns),
compared to ISGW2
\protect\cite{isgw2}, SISM \protect\cite{suzuki}, CNP
\protect\cite{cnp}, and SHJL \protect\cite{holdom} results. }
\label{tab3}
\begin{center}
\begin{tabular}{|c||c|c|c||c|c|c|c|}
\hline
\hline
Doublet&
RFTC& DESC &SEVC &ISGW2 &SISM&CNP&SHJL \\
\hline
 $C/C^{*}$
& 0.35 & 0.36 & 0.35 &0.48 &0.34 &0.33 &0.31\\
$E/E^{*}$
& 0.71& 0.72 & 0.71 &1.00 & 0.76 & 0.71&0.70 \\
$F/F^{*}$
& 0.60 & 0.63 & 0.60 &2.00 & 0.70 & 0.50 & 1.54 \\
$G/G^{*}$
 & 1.00 & 1.00 &1.00 &not given & 0.98 &not given & 7.00\\
$C_{2}/C_{2}^{*}$
 & 0.49 & 0.50 & 0.49 &0.00 & 0.41&not given &not given\\
 $E_{2}/E^{*}_{2}$
&0.82 &0.72 & 0.79 &not given & not given &not given&not given\\
 $F_{2}/F^{*}_{2}$
& 0.67& 0.64 & 0.66 &not given & not given&not given&not given\\
\hline
\hline
\end{tabular}
\end{center}
\end{table}


\newpage

\begin{thebibliography}{99}
\bibitem{isgur} N. Isgur and M. B. Wise,
Phys. Lett. B {\bf 232}, 113 (1989);
Phys. Lett. B {\bf 237}, 527 (1990).
\bibitem{reviews} There are many excellent reviews on
the subject of  heavy
quark symmetry and effective theory. See, for
example,
H. Georgi, in {\em Perspectives in the Standard Model},
Proceedings of the Theoretical Advanced Study Institute,
Boulder, Colorado, 1991, edited by R. K. Ellis et al.,
World Scientific, Singapore, 1991;
N. Isgur and M. B. Wise, in {\em Heavy Flavors}, edited by. A. J. Buras
and M. Lindner, World Scientific, Singapore, 1992;
M. Neubert,
Phys. Rept. {\bf 245}, 259 (1994).
\bibitem{modelling} S. Veseli and M. G. Olsson,
Phys. Lett. B {\bf 367}, 302 (1996).
\bibitem{cleo} J. P. Alexander et al., {\em Semi-leptonic
$B$ meson decay to $P$ wave charm mesons}, report CLEO-CONF-95-30.
\bibitem{aleph} D. Buskulic et al.,
Phys. Lett. B {\bf 345}, 103 (1995).
\bibitem{opal} R. Akers et al., OPAL Collaboration,
Z. Phys. C {\bf 67}, 57 (1995).
\bibitem{bjsum} J. D. Bjorken,
{\em New symmetries in heavy flavor physics},
in {\em Results and Perspectives
in Particle Physics}, Proceedings
of the $4^{\rm th}$ Recontres de Physique
de la Vall$\acute{\rm e}$e D'Aoste, La Thuile, Italy, 1990, edited
by A. Greco, Editions Fronti$\grave{\rm e}$res, Gif-sur-Yvette, France,
1990.
\bibitem{bjsum2} N. Isgur and M. B. Wise,
Phys. Rev. D {\bf 43}, 819 (1991).
\bibitem{voloshin} M. B. Voloshin,
Phys. Rev. D {\bf 46}, 3062 (1992).
\bibitem{georgi2} H. Georgi,
Nucl. Phys. B {\bf 348}, 293 (1991).
\bibitem{korner} J. G. K$\ddot{\rm o}$rner and G. A. Schuler,
Z. Phys. C {\bf 38}, 511 (1988).
\bibitem{falk} A. F. Falk,
Nucl. Phys. B {\bf 378}, 79 (1992).
\bibitem{mannel} A. Ali, T. Ohl, and T. Mannel,
Phys. Lett. B {\bf 298}, 195 (1993).
\bibitem{neubert} M. Neubert,
Phys. Lett B {\bf 264}, 455 (1991).
\bibitem{sadzikowski} M. Sadzikowski and K. Zalewski,
Z. Phys. C {\bf 59}, 677 (1993).
\bibitem{suzuki} T. B. Suzuki, T. Ito, S. Sawada, and M. Matsuda,
Prog. Theor. Phys. {\bf 91}, 757 (1994).
\bibitem{stop} S. Veseli and M. G. Olsson, {\em
S to P Wave Form Factors in Semi-leptonic B Decays}, UW-Madison report
MADPH-95-907 (hep-ph/9509230), to appear in Z. Phys. C.
\bibitem{zalewski} K. Zalewski,
Phys. Lett. B {\bf 264}, 432 (1991).
\bibitem{observations} M. G. Olsson, S. Veseli, and K. Williams,
Phys. Rev. D {\bf 51}, 5079 (1995).
\bibitem{fermionic} M. G. Olsson, S. Veseli, and K. Williams,
Phys. Rev. D {\bf 53}, 4006 (1996).
\bibitem{pdg} L. Montanet et al., Particle Data Group,
Phys. Rev. D {\bf 50}, 1173 (1994).
\bibitem{goebel} C. Goebel, D. LaCourse, and M. G. Olsson,
Phys. Rev. D {\bf 41}, 2917 (1990).
\bibitem{barger} V. D. Barger and D. B. Cline, {\em
Phenomenological Theories of High Energy Scattering},
W. A. Benjamin Inc., New York, 1969.
\bibitem{bethe} E. E. Salpeter and H. A. Bethe,
Phys. Rev. {\bf 76}, 1232 (1949).
\bibitem{gellmann} M. Gell-Mann and F. Low,
Phys. Rev. {\bf 84}, 350 (1951).
\bibitem{salpeter} E. E. Salpeter,
Phys. Rev. {\bf 87}, 328 (1952).
\bibitem{long} C. Long and D. Robson,
Phys. Rev. D {\bf 27}, 644 (1983).
\bibitem{sucher} G. Hardekopf and J. Sucher,
Phys. Rev. A {\bf 30}, 703 (1984);
Phys. Rev A {\bf 31}, 2020 (1985).
\bibitem{plesset} Milton S. Plesset,
Phys. Rev. {\bf 41}, 278 (1932).
\bibitem{lagae} J.-F. Laga$\ddot{\rm e}$,
Phys. Rev. D {\bf 45}, 317 (1992).
\bibitem{instantaneous} M. G. Olsson, S. Veseli, and
K. Williams, Phys. Rev. D {\bf 52}, 5141 (1995).
\bibitem{lacourse} D. LaCourse and M. G. Olsson,
Phys. Rev. D {\bf 39}, 2751 (1989).
\bibitem{aft} M. G. Olsson and S. Veseli,
Phys. Rev. D {\bf 51}, 3578 (1995).
\bibitem{argus} H. Albrecht et al., ARGUS Collaboration,
Z. Phys. C {\bf 57}, 533 (1993).
\bibitem{cleo2} B. Barish et al., CLEO Collaboration,
Phys. Rev. D {\bf 51}, 1014 (1995).
\bibitem{isgw2} D. Scora and N. Isgur,
Phys. Rev. D {\bf 52}, 2783 (1995).
\bibitem{cnp} P. Colangelo, G. Nardulli, and N. Paver,
report BARI TH/93-132, UTS-UFT-93-3 (hep-ph/9303220);
Phys. Lett. B {\bf 293}, 207 (1992).
\bibitem{holdom} M. Sutherland, B. Holdom, S. Jaimungal, and R. Lewis,
Phys. Rev. D {\bf 51}, 5053 (1995).
\bibitem{cleo3} Y. Kubota et al., {\em Measurement of
the $B(B\rightarrow D_{0}e^{-}\bar{\nu}$) using neutrino
reconstruction techniques}, report CLEO-CONF-95-10.
\bibitem{isgurp}N. Isgur, private communication.
\bibitem{altarelli} G. Altarelli and S. Petrarca,
Phys. Lett. B {\bf 261}, 303 (1991).
\bibitem{bigi} I. Bigi, R. Blok, M. Shifman, and A. Vainshtein,
Phys. Lett. B {\bf 323}, 408 (1994).
\bibitem{cortes} J. L. Cortes, X. Y. Pham, and A. Tounsi,
Phys. Rev. D {\bf 25}, 188 (1982).
\bibitem{gronberg} J. Gronberg et al., {\em Measurement
of the branching ratio $B^{-}\rightarrow Xe^{-} \bar{\nu}_{e}$ with lepton
tags}, report CLEO-CONF-94-6.
\bibitem{rftiwf} M. G. Olsson and S. Veseli,
Phys. Rev. D {\bf 51}, 2224 (1995).
\end{thebibliography}
\end{document}